\documentclass[iop,revtex4]{emulateapj}
\usepackage{lineno}

\begin{document}

\renewcommand{\topfraction}{1.0}
\renewcommand{\bottomfraction}{1.0}
\renewcommand{\textfraction}{0.0}

\title{New orbits based on speckle interferometry at SOAR$^1$}

\altaffiltext{1}{Based on observations obtained  at the Southern Astrophysical Research
(SOAR) telescope. }

\author{Andrei Tokovinin}
\affil{Cerro Tololo Inter-American Observatory, Casilla 603, La Serena, Chile}
\email{atokovinin@ctio.noao.edu}

\begin{abstract}
Orbits of  55 visual  binary stars are  computed using  recent speckle
interferometry data from the  SOAR telescope: 33 first-time orbits and
22 revisions of previous orbit calculations. The orbital periods range
from 1.4 to 370 years, the quality of orbits ranges from definitive to
preliminary and  tentative.  Most binaries consist  of low-mass dwarfs
and  have  short periods  (median  period  31  years).  The  dynamical
parallaxes  and  masses  are   evaluated  and  compared  to  the  {\it
  Hipparcos} parallaxes. Using differential speckle photometry, binary
components are placed on the color-magnitude diagram.
\end{abstract} 

\maketitle

\section{Introduction}
\label{sec:intro}

This paper  presents new  or updated orbits  for 55 binary  systems or
subsystems. It  is based on speckle  interferometric measurements made
at  the   4.1-m  Southern  Astrophyisical   Research  (SOAR)  telescope
\citep{TMH10,SAM09,Tok2012a,TMH14,TMH15,SAM2015}     combined     with
archival  data collected in  the Washington  Double Star  Catalog, WDS
\citep{WDS}. It continues previous work on binary orbits resulting from
the SOAR speckle program. 

The   Sixth   Catalog   of   Visual   Binary   Orbits,   VB6\footnote{
  \url{http://www.usno.navy.mil/USNO/astrometry/optical-IR-prod/wds/orb6}
} \citep{VB6},  presently contains orbital elements of  more than 2600
pairs.  Knowledge of binary orbits is needed in various areas such as:
(i) measurement of stellar masses (especially for stars of very low or
high  mass, unusual  chemical composition  or at  various evolutionary
stages); (ii) statistics of orbital elements in relation to mechanisms
of binary  formation and evolution;  (iii) accurate models  of stellar
motion for  astrometry; (iv) binary  systems of special  interest, for
example exohosts  or young binaries with  circumstellar material.  The
latter class is poorly  defined, given that a common ``uninteresting''
binary may suddenly become important  in the light of new discoveries,
e.g.  the  visual   triple  star   HD~131399   hosting   an  unusual   planet
\citep{Wagner2016}.   Accurate parallaxes  soon to  be  available from
 {\it Gaia} will  greatly enhance the value of  binary orbits for
 calibration  of  stellar  masses.   During five  years  of  this
  mission, only  portions of long binary periods  will be covered,
 making  the  ground-based  monitoring an  essential  complement.
 These arguments indicate that the  work on binary orbits is part
 of the astronomical infrastructure and has its own value.

Most binaries  studied here consist  of late-type dwarfs in  the solar
neighborhood,   including   close  pairs   first   resolved  by   {\it
  Hipparcos}. These  pairs have orbital periods measured  in years and
decades, while  wider classical visual binaries have  periods of a
few  centuries or  millenia. Efforts  to monitor  the motion  of these
fast  binaries  by  speckle   interferometry  have  been  made  by
\citet{Balega2006},  \citet{Horch2015},  and   others  mostly  on  the
northern sky. The SOAR data on their southern counterparts reveal a fast
motion and the  lack of coverage after the  {\it Hipparcos} discovery in
1991.   First, still  preliminary  orbits of  several  such pairs  are
presented here.

Another class  of preliminary orbits have long  periods and incomplete
orbit coverage, so typical  for visual orbits.  Unlike spectroscopists
who  accumulate data for  one or  several orbital  revolutions before
publishing the orbit,  the visual binary community has  a tradition of
sharing  the  observations,  because  the  time needed  for  an  orbit
calculation  may exceed  the human  life  span. The  downside of  this
tradition is  publication of premature orbits  and frequent, sometimes
poorly justified  orbit revisions.  One of the  reasons for publishing
here uncertain first-time orbits is the need to understand the orbital
motion and to plan further observations for their improvement.

Compared to  visual micrometer measures, speckle  interferometry has a
much  improved accuracy  (the  SOAR data  have  random and  systematic
errors of a few mas).  Some classical binaries discovered and measured
at  the limit  of visual  techniques  have now  reliable and  accurate
orbits based on speckle  interferometry.  Large resolving power of the
4.1-m SOAR telescope allows to map previously inaccessible portions of
their  orbits near  periastron.   This is  particularly important  for
difficult orbits with high eccentricity or oriented edge-on.  Such
orbits  are  under-represented  in  the  VB6  and  hence  distort  the
statistics. 

Traditionally,  visual  binary  orbits  were mostly  computed  by  the
observers themselves who had a deep understanding of the underlying data
and  its reliability.   Speckle  interferometry is  more reliable  than
micrometer measures; however, measures  of close or very unequal pairs
near the limit of the technique can also be unreliable or distorted by
instrumental  artifacts.  Access  to  the original  SOAR  data and  the
possibility  to reprocess  doubtful measures  distinguishes  this work
from orbit calculation made by others.

\begin{figure*}
\epsscale{1.0}
\plotone{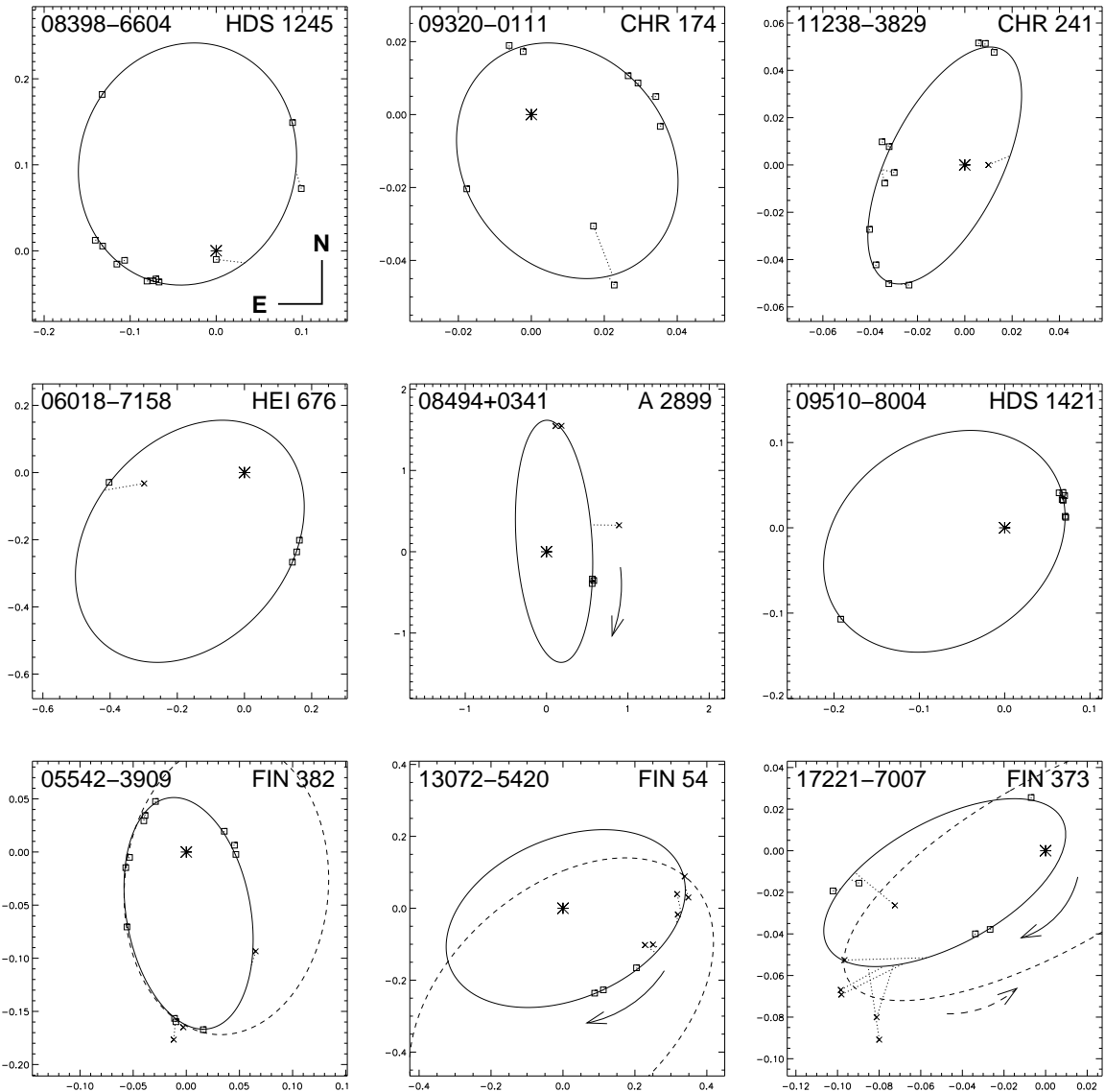}
\caption{Plots  of selected  orbits.   Top row:  reliable new  orbits,
  middle row: preliminary new orbits, bottom row: orbit revisions (old
  orbits  in dashed  line).  In  each plot,  the primary  component is
  located  at the coordinate  origin  and  the scale  is  in  arcseconds.
  Squares  and   crosses  denote  speckle   and  visual  measurements,
  respectively.   They  are connected  by  dotted  lines to the  ephemeris
  positions on the orbital ellipse (full line).
\label{fig:mosaic} }
\end{figure*}


\section{Orbital elements}
\label{sec:orb}

Orbital  elements and  their  errors are  determined by  least-squares
fitting     using    the     IDL    code     {\tt    ORBIT}.\footnote{
  \url{http://www.ctio.noao.edu/\~{}atokovin/orbit/index.html}  } When
no  prior orbit  is  available, the  initial  approximation is  chosen
interactively, considering the observed motion of each pair. This step
is  non-trivial  because position  angles  are  determined by  speckle
interferometry only modulo  180\degr, allowing quadrant ``flips'', the
orbit coverage often has large  gaps, and some measurements are simply
erroneous.   Sometimes the  same data can  be represented  by  several very
different orbits.

The least-squares method implies that the measurement errors are known
and  normally  distributed;  in  such  case the  optimum  weights  are
inversely  proportional   to  the  squares  of   the  errors.   Visual
measurements do not match  this model, having random and/or systematic
errors  that  are  difficult  to  quantify.   The  errors  of  speckle
interferometry behave better, but  depend on the instrument, telescope
aperture,  magnitude difference,  wavelength, data  quality,  etc., so
there  are no  simple  prescriptions for  their  evaluation. Here  the
errors are assigned subjectively based on these considerations and are
corrected  iteratively  by  assigning  larger  errors  (hence  smaller
weights)  to  the outliers.  The   result  is  checked  by  computing  the
$\chi^2/N$ metric, where $N$ is  the number of degrees of freedom.  In
the ideal  case, $\chi^2/N \sim 1$.  The same error in  the radial and
tangential directions  is adopted for  each observation. This  is true
for  speckle  interferometry,  while  the  known  tendency  of  visual
observers  to  measure  angles  more accurately  than  separations  is
ignored here,  considering  low weight of  the visual data.   For many
pairs observed at  SOAR, the rms residuals from the  orbits are on the
order of  1\,mas.  The  weighting scheme adopted  in the VB6  does not
fully account  for the high accuracy  of SOAR data; as  a result, some
recent  orbits  \citep[e.g.][]{TMH15}   are  ``pulled  away''  by  
inaccurate old measures and leave systematic residuals to the SOAR data.

As   mentioned  above,   for  some   binaries  the   scarce  available
observations  do not fully  constrain their  orbital elements  and are
compatible with  a wide range of  orbits. In such cases,  the mass sum
calculated using the HIP2  parallax \citep{HIP2} provides a guidance
on selecting the  most plausible orbit.  Some orbital  elements can be
fixed while fitting the remaining elements, so that the resulting mass
sum  takes  a reasonable  value.   This  approach  does not  work  for
binaries with unknown or small parallax.

Table~\ref{tab:orb}  lists the  orbital elements  and their  errors in
common notation  ($P$ -- orbital  period, $T_0$ -- epoch  of periastron,
$e$ -- eccentricity, $a$ -- semimajor axis, $\Omega$ -- position angle
of the  node for the equinox  J2000.0, $\omega$ -- argument  of periastron,
$i$ -- inclination). The first column gives the WDS code of the binary
and, in the following line, its {\it Hipparcos} number when available.
The  system identifier  adopted in  the WDS  (``discoverer  code'') is
given in  the second  column. For each  pair, the first  line contains
orbital elements, while the  following line gives their formal errors.
The errors  are omitted  for some preliminary  orbits of grade  5 where
they have  no sense.   Provisional grades are  assigned 
based on  the principles described  in the VB6,  where grades 4  and 5
mean  preliminary  orbits and  grade  1  are  definitive and  accurate
orbits.  The  last column contains references to   previously computed
orbits, when available.

Figure~\ref{fig:mosaic} gives  samples of orbital plots, illustrating
their different quality. Its top row shows three first-time,
but well defined orbits. The second row shows new preliminary
 orbits that risk substantial revisions in the future, mostly because
 the coverage is still insufficient. Drastic revisions of previous
 orbits are shown in the bottom row. 

Individual    observations    and     residuals    are    listed    in
Table~\ref{tab:obs},  available in  full  electronically. It  contains
still unpublished measures made at  SOAR in 2016, while some published
SOAR measures were reprocessed. Its first column identifies the binary
by its WDS code. Then follow  the time of observation $T$ in Besselian
years, position angle $\theta$ for the time of observation (correction
for   precession  is  done   internally  during   orbit  calculation),
separation  $\rho$  in  arcseconds,  and measurement  error  $\sigma$.
Unrealistically   large   errors  are   assigned   to  the   discarded
observations, so  that they have  no influence on the  fitted elements
but  are still  kept in  the Table.  
The  last two  columns  of Table~\ref{tab:obs}  contain the  residuals
O$-$C in angle and separation.

\begin{figure}
\epsscale{1.0}
\plotone{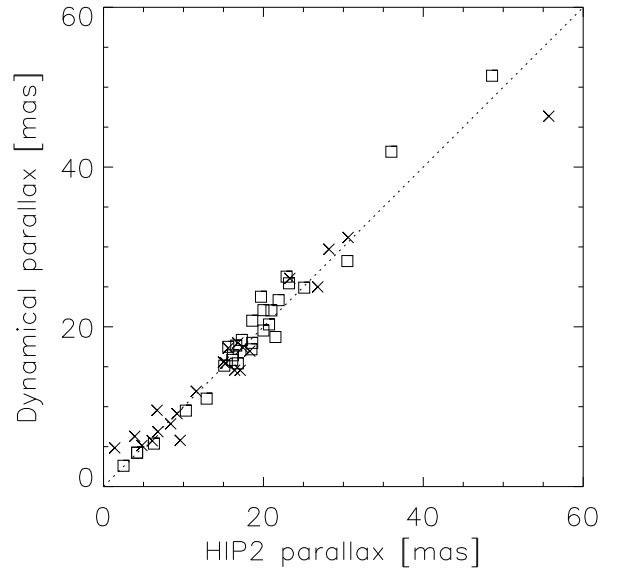}
\caption{Comparison between the dynamical and {\it Hipparcos} parallaxes.
Binaries with orbits of grade 3 or better are plotted by squares, the
remaining binaries by crosses.  
\label{fig:dynpar}}
\end{figure}

Table~\ref{tab:ptm}  provides additional  information, namely  the the
spectral  type as  given in  {\it Hipparcos}  or SIMBAD  and  the HIP2
parallax $\pi_{\rm  HIP2}$, to be  compared to the  dynamical parallax
$\pi_{\rm dyn}$  in the  next column. The  latter is evaluated  by the
Baize-Romani  method,  applicable when  binary  components follow  the
standard   main-sequence   relation    between   mass   and   absolute
magnitude. Taking the initial mass sum ${\cal M} = 2{\cal M}_\odot$, I
compute  the dynamical  parallax $\pi_{\rm  dyn} =  a  {\cal M}^{-1/3}
P^{-2/3}$,  use it  to evaluate  the absolute  $V$ magnitudes  of both
components, and estimate their masses ${\cal M}_1$ and ${\cal M}_2$ by
means of the standard relation  by \citet{HM93}, valid for ${\cal M} <
2{\cal M}_\odot$  and extrapolated  to larger masses  in a  few cases.
The new  mass sum leads to the  new value of $\pi_{\rm  dyn}$, and the
iterations converge  rapidly. The components' masses  ${\cal M}_1$ and
${\cal M}_2$ obtained in this  procedure are listed in the columns (6)
and  (7)   of  Table~\ref{tab:ptm}.   Asterisks   mark  the  dynamical
parallaxes  derived  from  reliable  orbits  of  grade  3  or  better.
Figure~\ref{fig:dynpar}   compares  the dynamical   and   {\it  Hipparcos}
parallaxes.

The combined  $V$ magnitude and $V  - I_C$ color index  in columns (8)
and  (9) are  taken  mostly  from the  {\it  Hipparcos} catalog.   For
HIP~83716B and 87914B, the  combined magnitudes refer to the secondary
subsystems,  while  $V -  I_C  =  1.0$ mag  is  assumed  to match  the
estimated spectral type  K0V of both pairs.  The  last four columns of
Table~\ref{tab:ptm} provide differential photometry resulting from the
SOAR  speckle  interferometry, where  the  filters  $y$  and $I$  have
central  wavelengths   and  bandwidths  of  543/22   and  788/132  nm,
respectively.  The $\Delta y$ and $\Delta I$ are average values, while
$\sigma_{\Delta y} $ and $\sigma_{\Delta I}$ stand for the rms scatter
of  magnitude difference  in each  filter if  measured  several times,
indicating  internal consistency of  the differential  photometry. For
some pairs  this scatter is as  good as 0.1 mag,  but it can  be up to
$\sim$0.5 mag.  Speckle differential photometry is unreliable for very
close  pairs at  or below  the diffraction  limit. For  faint  or wide
binaries,  $\Delta m$  can be  systematically  overestimated.  Despite
these caveats,  speckle interferometry at  SOAR is the only
source of differential photometry for some close binaries.

\begin{figure}
\epsscale{1.0}
\plotone{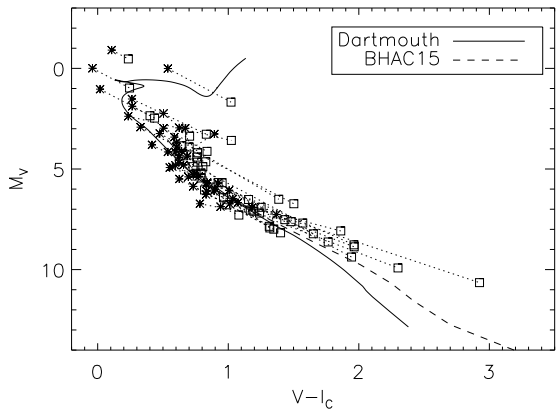}
\caption{Color-magnitude diagram. The primary and secondary components
  of each pair are plotted as asterisks and squares, respectively, and
  connected by dotted lines. The  full line is the Dartmouth isochrone
  \citep{Dotter2008},  the dashed  line is  the  BHAC15 \citep{BHAC15}
  isochrone, both for 1~Gyr age and solar metallicity.
\label{fig:cmd}  }
\end{figure}

When both $\Delta y$ and $\Delta I$ are measured, the empirical linear
relation $\Delta I \approx 0.7 \Delta y$ holds, allowing to estimate
differential magnitudes in both filters even when measurements in only
one filter are available.  It is also reasonable to assume that
$\Delta V \approx \Delta y$ and $\Delta I_C \approx \Delta
I$. Therefore, magnitudes and colors of each binary component can be
estimated using combined and differential photometry. 

Figure~\ref{fig:cmd} shows the  $(M_V, V-I_C)$ color-magnitude diagram
(CMD)  where  only  binaries   with  $\pi_{\rm  HIP2}  >  4$\,mas  are
plotted. For  reference, the 1-Gyr  isochrones from \citet{Dotter2008}
and \citet{BHAC15} are shown. 


\section{Notes on individual binaries}
\label{sec:notes}

Brief comments on some binaries are provided in this Section. The
acronym RV stands for radial velocity. 

{\it  04180$-$3826.} HDS~546  is a  typical case  of a  {\it Hipparcos}
binary that has  not been observed after its  discovery in 1991. It
has made one full revolution since, and the short arc observed at SOAR
shows a rapid  motion. This  is a pair  of K2V  chromospherically active
dwarfs.

{\it  05229-4219.}  TOK~93  Aa,Ab  was  first  resolved in  2011  at  Gemini-S
\citep{NICI}. It has a  variable RV and astrometric acceleration. Some
observations have been re-processed. The faint star B at 14\arcsec ~is
physical.

{\it 05542-2909.}   The orbit of  FIN~382 with $P=20.16$\,yr  given in
\citep{TMH15} is  wrong (Figure~\ref{fig:mosaic}), the  true period is
10\,yr. The revised orbit has a very high quality, with the rms residuals
of 3.6 and 1.8 mas in two coordinates.

{\it 08369$-$7857.}  KOH~79 is a  young variable star EG~Cha in  the $\eta$~Cha
association.  Two preliminary orbits  proposed by  \citet{Koh2002} are
now replaced  by one  non-ambiguous, though still  preliminary, orbit.
The dynamical parallax of 11.5\,mas matches the known distance to this
association and may eventually help to measure it better.

{\it 09149+0427.}  HEI~350 is a nearby red dwarf pair GJ~390 (the spectral
type of  the primary is K4V).   Despite the large $\Delta  m$, the two
first  micromemeter  measures  by  W.~Heintz  in 1987  and  1988  have
contradictory quadrants, and both do  not agree well with the accurate
measures available since 2003.  The pair has not been resolved by {\it
  Hiparcos}, presumably because of the  large $\Delta m$. The orbit is
only loosely constrained.  The binary might be young and belong to the
Castor moving group. Its secondary  with $V-I_C = 3.15$ mag is
located  above the main sequence in  Figure~\ref{fig:cmd}; the large magnitude
difference $\Delta V = 4.0$ mag has been confirmed in 2016.4. 


{\it  09275-5806.} The orbit of CHR~240 published in \citep{TMH15} is
substantially revised, with a new period of 1.42\,yr instead of
2.62\,yr. This  very close pair is often under the diffraction
limit of SOAR. 

{\it 09466$-$4966.}  Our relative  photometry implies that $\Delta I >
\Delta y$,  i.e. the  secondary is bluer  than the primary.   The G5IV
primary is evolved; it is located in the subgiant branch of the CMD.  The
{\it Hipparcos} photometry  $\Delta Hp = 1.24$ mag  is suspect because
in 1991 the pair was very close, 0\farcs15.

{\it 09510-8004.}  HDS~1421 (B8IV) belongs to the Sco-Cen association;
the dynamical  parallax is 6.4\,mas.  The orbit is  still preliminary,
but its curvature is well defined.

{\it 11102-1122.}  HDS~1590 (HD~97038) is a double-lined spectroscopic
binary (D.~Latham, 2012,  private communication). Its corrected orbit
is now well defined after coverage of the periastron.

{\it  13072$-$5420.}  Despite  the  drastic revision  of the  previous
orbit, the elements of FIN~54 remain  tentative, given that only a
small arc is covered  (Figure~\ref{fig:mosaic}). One of the components
is an eclipsing binary  V949~Cen.  The dynamical parallax is 4.8\,mas,
substantially larger than 1.3\,mas measured by {\it Hipparcos}. 

{\it  13377$-$2337.}  Two  orbits of  RST~2856  (BD$-$22\degr~3633) by
\citet{Hei1997} with periods of 110~yr and 51.5~yr are now replaced by
the new, but  still tentative orbit with $P=122$\,yr.   It is based on
two HRCam measures  and a handful of visual  micrometer measures.  The
$\Delta y = 2.1$ mag is derived from the very noisy data and is likely
wrong; the pair  has not been resolved by  {\it Hipparcos} despite its
0\farcs3  separation.   The   dynamical  parallax  is  12.3\,mas,  the
matching spectral type of the primary is G5V.

{\it  13598$-$0333.} This  star  HR~5278 (F6V)  possibly contains  a
debris   disk.   A   slight  revision   of  the   previous   orbit  by
\citet{Hor2011b} leaves  very well defined  elements.  Both components
are at  about 1.5 mag above the  main sequence, while the  mass sum is
slightly larger than  expected.  The stars could have  evolved off the
main  sequence  or are  still  contracting.  

{\it  14383$-$4954.} A drastic revision of the orbit by \citet{Zir2014a}
changes the period of FIN~371 from 27.6\,yr to 50.7\,yr.   

{\it  14453$-$3609.} The first  tentative  orbit for  I~528,  with a  poor
coverage  and  doubtful visual  measures, is suggested.  The  formal  errors of  the
elements are misleadingly small.  The component C is at 4\farcs5.  The
dynamic parallax is 5.3\,mas, and the binary may belong to the Sco-Cen
association.

{\it  14589+0636.} WSI~81  (HIP  73314) is  on  the California  planet
search program  \citep{Isaacson2010}.  The  new orbit is  well defined
and leads  to the mass sum  of 2.2 ${\cal  M}_\odot$, while the
dynamical parallax corresponds to 1.7 ${\cal  M}_\odot$.  Both components
are located  on the main  sequence. 


{\it  15035$-$4035.}  The revision  of the  orbit by  \citet{Hei1981a} for
I~1262 nearly  doubles its period, from  134 to 251 yr,  but the orbit
still remains  preliminary and poorly constrained.  The binary belongs
to the Sco-Cen association, its dynamic parallax is 5.7\,mas.

{\it 15042$-$1530.} The new edge-on ($i=89\degr$) orbit of RST~3906 is
very  different  from  the   previous  one  by  \citet{Hei1981a}  with
$i=129\degr$. The  new orbit  fits the speckle  data well  but implies
that many micrometer measures of this pair are totally wrong (possibly
observers confused it  with another pair?). The pair was unresolved at
SOAR on 2016.14 in agreement with the new orbit.

{\it  15282$-$0921.} This nearby  K2V star  GJ~586 is  a spectroscopic
binary with a record-high eccentricity of $e=0.976$.  Precise RVs from
\citet{Stressmeier2013}  constrain  all  spectroscopic  elements  very
tightly,  leaving for  adjustment  only the  ``visual'' elements  $a$,
$\Omega$, and $i$.   The orbital parallax is $52.5  \pm 6.2$\,mas, the
orbital  masses  are  $0.86  \pm  0.11$ and  $0.58  \pm  0.08$  ${\cal
  M}_\odot$.

{\it  15290$-$2852.}   BU~1114  is  a  nearby dwarf  with  a  physical
tertiary component  C at 9\farcs6.  Some  published non-resolutions of
AB  at SOAR refer  in fact  to this  component C,  pointed erroneously
instead of  AB.  The edge-on orbit is  preliminary and yields the  mass sum of
2.0 ${\cal  M}_\odot$, less than  2.9 ${\cal M}_\odot$  estimated from
the luminosity. 

{\it  15317+0053.}  TOK~48 (HIP  76031, HD  138369) is  a single-lined
spectroscopic    binary    with     a    period    of    619.3    days
\citep{Griffin2013}. Resolved measures from SOAR are combined with the
RVs  of  the primary  component  in  the  orbital fit.   However,  the
solution converges to $i \approx 180\degr$, contradicting the observed
RV variation.  Therefore,  in the combined speckle-spectroscopic orbit
the inclination  was fixed  at 150\degr.  The  differential photometry
$\Delta y =  1.6$ mag and $\Delta  I = 1.25$ mag leaves  no doubt that
the spectrum  should contain double  lines. However, Griffin  could no
find  them,  despite  dedicated   effort.   His  hypothesis  that  the
secondary component is a close  binary is not supported by the speckle
photometry,  but  the secondary  could  have  a  fast axial  rotation,
reducing   the depth  of its  lines in  the blended  spectrum. Both
components are located on the main sequence.

{\it  15332$-$2429.}  CHR~232  Aa,Ab (HR  5765) is  an  A7V chemically
peculiar  binary \citep{Abt1995} with  a new  well-constrained 16.5-yr
orbit (rms  residuals 1\,mas). A very similar  orbit has just been
  published by \citet{Doc2016}.   It has  made  more  than one  full
revolution since its discovery in  1996.18.  The visual companion B at
9\farcs2 is  also a  resolved close binary  SEE~238 BC with  a 60.4-yr
orbit; its  observation with  HRCam in 2008.54  in fact refers  to the
Aa,Ab pair.  The two subsystems Aa,Ab and BC in this 2+2 quadruple are
not co-planar.  The  dynamical parallaxes of Aa,Ab and  BC are 9.4 and
10.3  mas, respectively, while  the HIP2  parallax is  10.4\,mas.  The
component Ab is located above the main sequence in the CMD.
  
{\it  15348$-$2808.}  The first  preliminary 26-yr  orbit of  TOK~49  Aa,Ab is
computed. The outer binary RST~1847 AB has a separation of 1\arcsec.

{\it  15548$-$6554.} The  first orbit  of the nearby  red dwarf  binary NZO~65
(G2V) is well constrained. Its components are on the main sequence.

{\it 16544$-$3806.} This is a  dramatic revision of the previous orbit
of HDS~2392 by \citet{TMH15}.

{\it 17018$-$5108.} The orbit of I~1306 by \citet{Ole2004b} with $P=62.3$~yr is
revised to $P=194$~yr  and still remains preliminary; only a half of it
is covered in almost a century  of observation.  This is a bright star
HR~6312, A9III.  The  mass sum of 1.0 ${\cal  M}_\odot$ does not match
the spectral type, throwing suspicion on the HIP2 parallax of 9.6\,mas
(the dynamical parallax is 5.8\,mas).

{\it  17066+0039.} After a slight correction, the orbit of the subsystem Ba,Bb
with $P=6.34$\,yr is now well constrained. The outer pair BU~823~AB
has a poorly defined 532-yr orbit. Coplanarity between the inner and
outer orbits is not excluded. 

{\it  17221$-$7007.} A  drastic  revision   of  the  orbit  of  FIN~373
(HR~6411, $i$~Aps, B8/9V). Compared  to the orbit by \citet{Doc2013d},
even  the direction  of  the  rotation has  changed  from prograde  to
retrograde, see Figure~\ref{fig:mosaic}.  The published SOAR measure in
2009.26 was  wrong and  corresponded to the  image doubling  caused by
telescope vibration;  it has been  reprocessed.  The rms  residuals to
the new orbit are 1.8\,mas, and it is now well defined.

{\it  17575$-$5740.}  The first  9-yr orbit  of  the subsystem  Ba,Bb is  well
constrained. The  outer pair  AB has a  separation of 2\farcs5  and no
orbit so far.

{\it 17584+0428.} KUI 84 (GJ 9609, K8) is a nearby close triple system
where the secondary component of  the 14.7-yr visual binary contains a
34.5-day  spectroscopic   subsystem  \citep{Tok1994}.   The  secondary
component is located above the  main sequence in the CMD, as expected.
Deviations of  the speckle  measures from the  latest visual  orbit by
\citet{Doc2005f}  prompted its  slight  revision here.  RVs were  used
together  with the speckle  measures, leading  to a  very well-defined
orbit. The HIP2  parallax of 22.9\,mas corresponds to  the mass sum of
2.0 ${\cal M}_\odot$, while the  orbital parallax of 26.3\,mas and the
mass sum  of 1.3  ${\cal M}_\odot$  are a better  match to  the masses
estimated by \citet{Tok1994}.

{\it 18150$-$5018.}   The first orbit of I~429  (HD~166839, A0V) looks
reasonable, but  then the  discovery measure by  Innes in 1902.5  at a
separation of 0\farcs5 must be  totally wrong. This and other measures
made before 1929 were ignored in the orbit calculation.  Presently the
pair is approaching  periastron. 

{\it  18368$-$2617.} The  new  orbit of  RST~2187~AB  looks good,  but
corresponds to the mass sum of 1.4 ${\cal M}_\odot$. A revision of the
HIP2 parallax from 21.5$\pm$1.7\,mas  to 19\,mas is needed to increase
the  mass sum  of this  pair of  nearly equal  G3V stars  to  2 ${\cal
  M}_\odot$.  The tertiary component C at 12\farcs6 is physical.

{\it  18480$-$1009.}  HDS~2665  is  a nearby  K2V  dwarf. The  primary
component  is located  on the  main sequence,  while the  secondary is
above it.   The mass  sum of 2.0  ${\cal M}_\odot$ is  somewhat larger
than expected, so there may be inner subsystems. 

{\it  19581$-$4808.} Some observations of the nearby G0V dwarf binary HDS~2842
were reprocessed for a minor revision of its previous 32-yr orbit, now definitive.

{\it 20217$-$3637.} HDS~2908 (HD~193464,  F8V) is another nearby dwarf
binary.   Double lines  were noticed,  but there  is  no spectroscopic
orbit in  the literature.  The $\Delta y  = 0.49$ mag is  based on two
well-resolved  observations and  is more  reliable than  $\Delta  Hp =
0.21$ mag.


{\it 22116$-$3428.} The 66-yr  first orbit of CHR~230~Aa,Ab (K1/K2III)
is well defined. The outer binary AB has a separation of 0\farcs9. The
large  eccentricity   $e=0.93$  of  Aa,Ab  could  be   caused  by  the
Kozai-Lidov cycles.   Both components are located on  the giant branch
in the CMD (Figure~\ref{fig:cmd}). 

\section{Summary}
\label{sec:sum}

New and updated orbits  presented here make an incremental improvement
of the VB6 content and thus  contribute, albeit in a small way, to the
observational foundations  of astronomy.   Some of these  binaries are
interesting because they are members of multiple systems with three or
more components, are young, lead to the useful measurements of masses,
or for other reasons.


\acknowledgments 


This work  used the  SIMBAD service operated  by Centre  des Donn\'ees
Stellaires  (Strasbourg, France),  bibliographic  references from  the
Astrophysics Data  System maintained  by SAO/NASA, and  the Washington
Double Star Catalog maintained at USNO.

{\it Facilities:}  \facility{SOAR}.



\clearpage

\LongTables

\begin{deluxetable}{c l rrr rrr r ccll}

\tabletypesize{\scriptsize}
\tablewidth{0pt}
\tablecaption{Orbital Elements \label{tab:orb}}
\tablehead{
\colhead{WDS} & 
\colhead{Discoverer} &
\colhead{$P$} & 
\colhead{$T_0$} &
\colhead{$e$} & 
\colhead{$a$} & 
\colhead{$\Omega$} &
\colhead{$\omega$} &
\colhead{$i$} &
\colhead{Gr} &
\colhead{Orbit} \\
 \colhead{HIP}   &
\colhead{Designation} &
\colhead{(yr)} & 
\colhead{(yr)} &
\colhead{ } & 
\colhead{($''$)} & 
\colhead{(\degr)} &
\colhead{(\degr)} &
\colhead{(\degr)} &
\colhead{ } & 
\colhead{Reference} 
}
\startdata
04180$-$3826 & HDS 546 & 22.674 & 2014.439 & 0.422 & 0.2333 & 53.2 & 211.1 & 110.6 & 4 & New\tablenotemark{a} \\
 20048 &    & $\pm$0.218 & $\pm$0.087 & $\pm$0.028 & $\pm$0.0052 & $\pm$2.2 & $\pm$5.2 & fixed&     &  \\
05229$-$4219 & TOK 93 Aa,Ab & 5.680 & 2013.511 & 0.481 & 0.0594 & 225.6 & 92.6 & 59.6 & 3 & New \\
 25148 &    & $\pm$0.270 & $\pm$0.088 & $\pm$0.041 & $\pm$0.0033 & $\pm$4.1 & $\pm$3.4 & $\pm$3.0&     &  \\
05542$-$2909 & FIN 382 & 10.070 & 2015.397 & 0.537 & 0.1117 & 194.8 & 200.1 & 127.7 & 1 & Tok2015c\tablenotemark{b} \\
 27901 &    & $\pm$0.036 & $\pm$0.018 & $\pm$0.004 & $\pm$0.0014 & $\pm$1.2 & $\pm$1.6 & $\pm$1.0&     &  \\
06018$-$7158 & HEI 676 & 102.0 & 2006.39 & 0.657 & 0.3988 & 301.9 & 340.0 & 169.0 & 5 & New\tablenotemark{a,b}\\
 28573 &    & $\pm$12.5 & $\pm$0.18 & $\pm$0.035 & $\pm$0.0237 & $\pm$1.2 & fixed & fixed &     &  \\
06032$+$1922 & HDS 823 Aa,Ab & 100 & 2044.6 & 0.93 & 0.449 & 51.5 & 80.2 & 95.9 & 5 & New \\
 28671 &    &  \ldots & \ldots & \ldots & \ldots & \ldots & \ldots & \ldots &     &  \\
08369$-$7857 & KOH 79 AB & 77.1 & 1996.4 & 0.291 & 0.2441 & 182.0 & 320.7 & 106.2 & 4 & Koh2002 \\
  \ldots &    & $\pm$8.9 & $\pm$1.5 & $\pm$0.052 & $\pm$0.0207 & $\pm$1.3 & $\pm$11.3 & $\pm$1.5&     &  \\
08398$-$6604 & HDS 1245 & 33.84 & 2013.175 & 0.781 & 0.2118 & 299.6 & 97.3 & 129.5 & 2 & New\tablenotemark{b} \\
 42496 &    & $\pm$0.90 & $\pm$0.055 & $\pm$0.013 & $\pm$0.0059 & $\pm$1.8 & $\pm$1.7 & $\pm$2.6&     &  \\
08494$+$0341 & A 2899 & 300  & 2123 & 0.231 & 1.53 & 4.0 & 250.4 & 107.8 & 5 & New\tablenotemark{a,b} \\
 43304 &    & fixed & $\pm$51 & $\pm$0.059 & $\pm$0.19 & $\pm$5.1 & $\pm$39.9 & $\pm$3.6&     &  \\
09149$+$0427 & HEI 350 & 62.68 & 1984.83 & 0.20 & 0.810 & 102.3 & 263.6 & 136.0 & 5 & Tok2015c\tablenotemark{a} \\
 45383  &    & $\pm$2.88 & $\pm$1.60 & fixed & $\pm$0.037 & $\pm$2.1 & $\pm$10.6 & $\pm$3.0&     &  \\
09186$+$2049 & HO 43 & 358.6 & 1941.2 & 0.40 & 0.610 & 80.0 & 214.9 & 131.6 & 4 & Baz1989a \\
 45671 &    & $\pm$9.4 & $\pm$2.5 & fixed    & $\pm$0.013 & $\pm$3.6 & $\pm$6.7 & $\pm$2.3&     &  \\
09275$-$5806 & CHR 240 & 1.421 & 2015.376 & 0.391 & 0.0324 & 259.2 & 141.3 & 130.0 & 2 & Tok2015c \\
 46388 &    & $\pm$0.007 & $\pm$0.027 & $\pm$0.038 & $\pm$0.0009 & $\pm$5.3 & $\pm$8.6 & fixed&     &  \\
09320$-$0111 & CHR 174 & 6.09 & 2015.06 & 0.471 & 0.0342 & 29.6 & 351.1 & 159.4 & 2 & New\tablenotemark{b} \\
 46776 &    & $\pm$0.14 & $\pm$0.04 & $\pm$0.019 & $\pm$0.0022 & $\pm$32.5 & $\pm$30.6 & $\pm$12.1&     &  \\
09466$-$4955 & HDS 1414 & 60 & 2018.67 & 0.80 & 0.144 & 291.3 & 243.1 & 62.8 & 5 & New\tablenotemark{a} \\
 47969 &    & fixed & $\pm$2.76    & $\pm$0.35 & $\pm$0.128 & $\pm$10.6 & $\pm$30.9 & $\pm$34.1&     &  \\
09510$-$8004 & HDS 1421 & 63.52 & 2015.24 & 0.50 & 0.1583 & 142.3 & 135.2 & 34.5 & 4 & New\tablenotemark{a,b} \\
 48320 &    & $\pm$5.68 & $\pm$0.37 & fixed & $\pm$0.0057 & $\pm$15.8 & $\pm$20.9 & $\pm$7.0&     &  \\
10112$-$3245 & HDS 1469 & 30.54 & 2015.323 & 0.636 & 0.1482 & 206.9 & 156.0 & 134.8 & 4 & New\tablenotemark{a}  \\
 49883 &    & $\pm$3.92 & $\pm$0.474    & fixed & $\pm$0.0235 & $\pm$11.2 & $\pm$17.6 & $\pm$15.6&     &  \\
11102$-$1122 & HDS 1590 & 20.61 & 1995.616 & 0.783 & 0.1613 & 254.7 & 74.6 & 120.7 & 3 & Tok2015c \\
 54580 &    & $\pm$0.43 & $\pm$0.424 & $\pm$0.009 & $\pm$0.0044 & $\pm$1.6 & $\pm$0.8 & $\pm$1.0&     &  \\
11238$-$3829 & CHR 241 & 3.559 & 2013.264 & 0.322 & 0.0576 & 154.2 & 257.8 & 114.7 & 3 & New \\
 55628 &    & $\pm$0.022 & $\pm$0.078 & $\pm$0.029 & $\pm$0.0014 & $\pm$2.1 & $\pm$3.7 & $\pm$1.8&     &  \\
11425$+$2355 & COU 390 & 92.75 & 2007.86 & 0.786 & 0.5498 & 231.4 & 100.9 & 43.6 & 3 & New \\
 57112 &    & $\pm$5.57      & $\pm$0.28 & $\pm$0.014 & $\pm$0.0168 & $\pm$3.2 & $\pm$2.7 & $\pm$3.2&     &  \\
11525$-$1408 & HDS 1676 & 16.71 & 1999.60 & 0.681 & 0.1474 & 324.6 & 307.5 & 48.2 & 3 & Tok2014a \\
 57894 &    & $\pm$0.42 & $\pm$0.42 & $\pm$0.010 & $\pm$0.0030 & $\pm$2.0 & $\pm$2.9 & $\pm$1.7&     &  \\
13072$-$5420 & FIN 54 & 373.7 & 2189.08 & 0.01 & 0.3516 & 113.3 & 79.7 & 135.3 & 5 & Hrt2011d\tablenotemark{a,b} \\
 64025 &    &  \ldots & \ldots & \ldots & \ldots & \ldots & \ldots & \ldots &     &  \\
13081$-$7719 & HDS 1839 & 17.44 & 2014.11 & 0.168 & 0.2007 & 349.2 & 320.5 & 120.5 & 4 & New\tablenotemark{a} \\
 64091 &    & $\pm$1.03 & $\pm$0.84   & $\pm$0.104 & $\pm$0.0124 & $\pm$5.4 & $\pm$19.0 & $\pm$2.5&     &  \\
13306$-$4238 & HDS 1891 & 12.51 & 2014.76 & 0.192 & 0.185 & 99.9 & 273.2 & 162.6 & 4 & New\tablenotemark{a} \\
 65906 &    & $\pm$0.05 & $\pm$1.38 & $\pm$0.046 & $\pm$0.027 & $\pm$103.7 & $\pm$157.7 & $\pm$27.1&     &  \\
13377$-$2337 & RST~2856 & 122 & 2038.6 & 0.55 & 0.352 & 108.7 & 290.2 & 81.9 & 5 & Hei1997 \\
 \ldots &    & \ldots & \ldots & \ldots & \ldots & \ldots & \ldots & \ldots &     &  \\
13598$-$0333 & HDS 1962 & 9.496 & 2008.139 & 0.374 & 0.0734 & 205.6 & 49.5 & 49.9 & 3 & Hor2011b \\
 68380 &    & $\pm$0.071 & $\pm$0.060 & $\pm$0.032 & $\pm$0.0029 & $\pm$3.7 & $\pm$2.7 & $\pm$3.2&     &  \\
14020$-$2108 & WSI 79 & 22.41 & 2015.258 & 0.735 & 0.1967 & 137.2 & 146.6 & 140.5 & 3 & Tok2015c \\
 68552 &    & $\pm$2.00 & $\pm$0.022 & $\pm$0.015 & $\pm$0.0067 & $\pm$3.1 & $\pm$5.0 & $\pm$2.8&     &  \\
14383$-$4954 & FIN 371 & 51.13 & 2015.39 & 0.205 & 0.0998 & 234.4 & 38.9 & 109.1 & 3 & Zir2014a \\
 71577 &    & $\pm$3.79 & $\pm$1.50 & $\pm$0.028 & $\pm$0.0041 & $\pm$1.7 & $\pm$13.6 & $\pm$1.5&     &  \\
14453$-$3609 & I 528 AB & 28.92 & 2008.87 & 0.45 & 0.0748 & 88.8 & 276.4 & 56.2 & 5 & New\tablenotemark{a} \\
 72140 &    & $\pm$0.93 & $\pm$1.11 & fixed & $\pm$0.0065 & $\pm$7.2 & $\pm$5.6 & $\pm$5.1&     &  \\
14589$+$0636 & WSI 81 & 9.364 & 2006.981 & 0.145 & 0.1344 & 65.0 & 279.1 & 130.5 & 3 & New \\
 73314 &    & $\pm$0.049 & $\pm$0.086 & $\pm$0.008 & $\pm$0.0015 & $\pm$1.7 & $\pm$4.5 & $\pm$1.1&     &  \\
15035$-$4035 & I 1262 & 235 & 1920.6 & 0.257 & 0.3122 & 33.6 & 189.6 & 52.6 & 5 & Hei1996a \\
 73667 &    &  \ldots & \ldots & \ldots & \ldots & \ldots & \ldots & \ldots &     &  \\
15042$-$1530 & RST 3906 & 53.897 & 2018.396 & 0.740 & 0.1570 & 154.5 & 3.7 & 89.4 & 4 & Hei1981a \\
 73724 &    & $\pm$15.772 & $\pm$3.868 & $\pm$0.337 & $\pm$0.0278 & $\pm$1.3 & $\pm$32.1 & $\pm$1.5&     &  \\
15251$-$2340 & RST 2957 & 53.52 & 2020.795 & 0.850 & 0.2762 & 96.4 & 302.0 & 86.0 & 4 & New \\
 75478 &    & $\pm$6.57 & $\pm$0.722 & $\pm$0.086 & $\pm$0.0076 & $\pm$1.3 & fixed & $\pm$0.6&     &  \\
15282$-$0921 & BAG 25 Aa,Ab & 2.43623 & 2007.2603 & 0.976 & 0.1074 & 272.8 & 255.6 & 55.4 & 2 & New \\
 75722 &    & fixed & fixed & fixed & $\pm$0.0081 & $\pm$2.1 & fixed & $\pm$3.6&     &  \\
15290$-$2852 & BU 1114 AB & 205 & 2047.16 & 0.498 & 0.7219 & 138.3 & 359.5 & 90.6 & 4 & New \\
 75790 &    & fixed & $\pm$4.45 & $\pm$0.045 & $\pm$0.0272 & $\pm$0.5 & $\pm$5.4 & $\pm$0.3&     &  \\
15317$+$0053 & TOK 48 & 1.713 & 2016.470 & 0.435 & 0.0427 & 161.9 & 45.8 & 150.0 & 3 & New \\
 76031 &    & $\pm$0.002 & $\pm$0.010 & $\pm$0.017 & $\pm$0.0010 & $\pm$2.1 & $\pm$2.3 & fixed&     &  \\
15332$-$2429 & CHR 232 Aa,Ab & 16.468 & 2011.536 & 0.419 & 0.0933 & 10.4 & 321.6 & 123.9 & 2 & New \\
 76143 &    & $\pm$0.194 & $\pm$0.064 & $\pm$0.009 & $\pm$0.0019 & $\pm$1.9 & $\pm$2.6 & $\pm$0.9&     &  \\
15348$-$2808 & TOK 49 Aa,Ab & 26.27 & 2006.28 & 0.342 & 0.1814 & 344.1 & 131.9 & 56.0 & 5 & New\tablenotemark{a}\\
 76275 &    & \ldots & \ldots & \ldots & \ldots & \ldots & \ldots & \ldots &     &  \\
15548$-$6554 & NZO 65 & 120.60 & 2009.88 & 0.464 & 0.6831 & 122.2 & 219.0 & 85.7 & 3 & New \\
 77921 &    & $\pm$2.18 & $\pm$0.11 & $\pm$0.010 & $\pm$0.0085 & $\pm$0.4 & fixed & $\pm$0.2&     &  \\
16544$-$3806 & HDS 2392 & 26.48 & 2009.459 & 0.50 & 0.1858 & 101.3 & 154.8 & 134.7 & 4 & Tok2015c \\
 82709 &    & $\pm$0.67 & $\pm$0.086 & fixed & $\pm$0.0045 & $\pm$3.2 & $\pm$2.8 & $\pm$2.0&     &  \\
17018$-$5108 & I 1306 & 193.7 & 1989.59 & 0.187 & 0.317 & 12.8 & 149.0 & 84.6 & 5 & Ole2004b \\
 83321 &    &  \ldots & \ldots & \ldots & \ldots & \ldots & \ldots & \ldots &     &  \\
17066$+$0039 & TOK 52 Ba,Bb & 6.339 & 2012.979 & 0.386 & 0.0725 & 15.9 & 179.2 & 34.0 & 2 & Tok2014a \\
 83716 &    & $\pm$0.146 & $\pm$0.076 & $\pm$0.023 & $\pm$0.0027 & $\pm$10.7 & $\pm$13.6 & $\pm$5.5&     &  \\
17221$-$7007 & FIN 373 & 56.90 & 2011.48 & 0.839 & 0.0779 & 323.0 & 51.2 & 126.3 & 3 & Doc2013d\tablenotemark{b} \\
 84979 &    & $\pm$7.67 & $\pm$0.39 & $\pm$0.023 & $\pm$0.0106 & $\pm$4.2 & $\pm$5.0 & $\pm$8.1&     &  \\
17575$-$5740 & TOK 55 Ba,Bb & 8.92 & 2015.16 & 0.387 & 0.1225 & 345.7 & 342.3 & 104.1 & 2 & New \\
 87914 &    & $\pm$0.36 & $\pm$0.13 & $\pm$0.020 & $\pm$0.0026 & $\pm$1.3 & $\pm$8.7 & $\pm$0.8&     &  \\
17584$+$0428 & KUI 84 & 14.715 & 2001.496 & 0.496 & 0.1746 & 351.5 & 190.0 & 62.8 & 1 & Doc2005f \\
 87991 &    & $\pm$0.040 & $\pm$0.051 & $\pm$0.008 & $\pm$0.0014 & $\pm$0.8 & $\pm$1.8 & $\pm$1.2&     &  \\
18092$-$2211 & RST 3157 & 10.685 & 2015.031 & 0.409 & 0.1575 & 232.4 & 55.5 & 50.2 & 2 & Tok2015c \\
 88932 &    & $\pm$0.045 & $\pm$0.020 & $\pm$0.008 & $\pm$0.0019 & $\pm$1.3 & $\pm$1.3 & $\pm$1.2&     &  \\
18150$-$5018 & I 429 & 63.10 & 2021.4 & 0.84 & 0.150 & 132.6 & 228.2 & 84.0 & 4 & New \\
 \ldots &    & $\pm$5.75 & $\pm$3.7 & $\pm$0.16 & $\pm$0.031 & $\pm$2.9 & $\pm$14.0 & $\pm$2.1&     &  \\
18368$-$2617 & RST 3187 AB & 18.37 & 2017.30 & 0.364 & 0.1661 & 40.6 & 7.2 & 70.1 & 3 & New \\
 91253 &    & $\pm$0.18 & $\pm$0.16 & $\pm$0.022 & $\pm$0.0020 & $\pm$0.9 & $\pm$2.9 & $\pm$0.8&     &  \\
18480$-$1009 & HDS 2665 & 31.212 & 2020.007 & 0.705 & 0.4535 & 192.8 & 44.2 & 49.1 & 3 & New \\
 92250 &    & $\pm$0.003 & $\pm$0.194 & $\pm$0.014 & $\pm$0.0178 & $\pm$1.3 & $\pm$3.4 & $\pm$3.4&     &  \\
19581$-$4808 & HDS 2842 & 35.28 & 2016.036 & 0.738 & 0.2364 & 254.2 & 60.6 & 70.2 & 3 & Tok2015c \\
 98274 &    & $\pm$2.48 & $\pm$0.065 & $\pm$0.037 & $\pm$0.0140 & $\pm$2.4 & $\pm$4.1 & $\pm$2.2&     &  \\
20057$-$3743 & HDS 2865 & 50.0 & 2018.71 & 0.328 & 0.1778 & 4.2 & 131.7 & 131.5 & 4 & New\tablenotemark{a} \\
 98979 &    & fixed & $\pm$0.56 & $\pm$0.040 & $\pm$0.0043 & $\pm$3.7 & $\pm$6.6 & $\pm$4.2&     &  \\
20104$-$1923 & HDS 2873 & 90.0 & 2007.46 & 0.484 & 0.3843 & 35.2 & -0.3 & 94.0 & 5 & New \\
 99391 &    & fixed & $\pm$0.67 & $\pm$0.006 & $\pm$0.0087 & $\pm$0.9 & $\pm$4.3 & $\pm$0.4&     &  \\
20202$-$3435 & I 1416 & 20.16 & 2000.19 & 0.90 & 0.163 & 140.1 & 37.2 & 113.9 & 3 & B\_\_1961d \\
100266 &    & $\pm$0.31 & $\pm$1.24 & fixed    & $\pm$0.035 & $\pm$3.2 & $\pm$19.7 & $\pm$11.6&     &  \\
20217$-$3637 & HDS 2908 & 12.74 & 2007.44 & 0.518 & 0.1120 & 107.8 & 338.9 & 80.8 & 3 & New \\
100417 &    & $\pm$0.24 & $\pm$0.52 & $\pm$0.046 & $\pm$0.0065 & $\pm$0.6 & $\pm$10.7 & $\pm$2.7&     &  \\
20248$-$1943 & HDS 2919 & 65.0 & 2020.90 & 0.474 & 0.3243 & 269.7 & 311.6 & 60.9 & 5 & New\tablenotemark{a} \\
100685 &    & fixed & $\pm$0.75 & $\pm$0.062 & $\pm$0.0086 & $\pm$2.9 & $\pm$3.6 & $\pm$2.5&     &  \\
21073$-$5702 & HDS 3009 & 54.08 & 2013.30 & 0.311 & 0.3482 & 350.1 & 143.3 & 56.8 & 3 & New \\
104256 &    & $\pm$4.21 & $\pm$0.75 & $\pm$0.030 & $\pm$0.0050 & $\pm$1.0 & $\pm$8.3 & $\pm$1.9&     &  \\
22116$-$3428 & CHR 230 Aa,Ab & 66.5 & 2010.37 & 0.929 & 0.1156 & 318.7 & 305.0 & 70.5 & 3 & New \\
109561 &    & $\pm$11.0 & $\pm$0.39& $\pm$0.024 & $\pm$0.0168 & $\pm$3.5 & $\pm$7.5 & $\pm$4.1&     &  
\enddata
\tablenotetext{a}{Insufficient coverage.}
\tablenotetext{b}{See Figure~1.}
\tablerefs{ 
B\_\_1961d --- \citet{B__1961d}; 
Baz1989a --- \citet{Baz1989a};
Doc2005f --- \citet{Doc2005f};
Doc2013d --- \citet{Doc2013d};
Hei1981a --- \citet{Hei1981a};
Hei1992b -- \citet{Hei1992b};
Hei1996a --- \citet{Hei1996a};
Hei1997 ---  \citet{Hei1997};
Hor2011b --- \citet{Hor2011b};
Hrt2011d --- \citet{Hrt2011d};
Koh2002 ---  \citet{Koh2002};
Ole2004b --- \citet{Ole2004b};
Tok2014a --- \citet{TMH14};
Tok2015c --- \citet{TMH15};
Zir2014a --- \citet{Zir2014a};
}
\end{deluxetable}


\begin{deluxetable}{c r rrr rr }
\tabletypesize{\scriptsize}
\tablewidth{0pt}
\tablecaption{Observations and residuals (Fragment) \label{tab:obs}}
\tablehead{
\colhead{WDS} & 
\colhead{$T$} &
\colhead{$\theta$} & 
\colhead{$\rho$} &
\colhead{$\sigma$} & 
\colhead{O$-$C$_\theta$} & 
\colhead{O$-$C$_\rho$} \\
& \colhead{(yr)} & 
\colhead{(\degr)} &
\colhead{($''$)} & 
\colhead{($''$)} & 
\colhead{(\degr)} &
\colhead{($''$)} 
}
\startdata
04180-3826 &  1991.2500 &  230.0 &   0.1360 &   0.0100 &    0.1 &  $-$0.0002 \\
04180-3826 &  2014.0430 &  228.6 &   0.1332 &   0.0015 &    0.3 &   0.0003 \\
04180-3826 &  2014.8536 &  207.0 &   0.1022 &   9.0047 &   $-$4.5 &   0.0051 \\
04180-3826 &  2015.1053 &  201.1 &   0.0826 &   0.0020 &   $-$1.8 &  $-$0.0006 \\
04180-3826 &  2015.7385 &  165.3 &   0.0566 &   0.0020 &    1.9 &   0.0001 \\
04180-3826 &  2015.9080 &  148.8 &   0.0559 &   0.0020 &    0.6 &   0.0008 \\
04180-3826 &  2016.1373 &  127.1 &   0.0578 &   0.0020 &   $-$1.3 &  $-$0.0010 
\enddata
\end{deluxetable}

\begin{deluxetable}{ccc cc cc rr rr rr}
\tabletypesize{\scriptsize}
\tablewidth{0pt}
\tablecaption{Parallaxes and photometry \label{tab:ptm}}
\tablehead{
\colhead{WDS} &
\colhead{HIP} &
\colhead{Spectral} & 
\colhead{${\pi_{\rm HIP2}}$} & 
\colhead{${\pi_{\rm dyn}}$} & 
\colhead{${\cal M}_1$} & 
\colhead{${\cal M}_2$} & 
\colhead{$V$} &
\colhead{$V-I_C$} &
\colhead{$\Delta y$} &
\colhead{$\sigma_{\Delta y}$} &
\colhead{$\Delta I$} &
\colhead{$\sigma_{\Delta I}$} \\
&  & 
\colhead{Type} &  
\colhead{(mas)} & 
\colhead{(mas)} & 
\colhead{(${\cal M}_\odot$)} & 
\colhead{(${\cal M}_\odot$)} & 
\colhead{(mag)} & 
\colhead{(mag)} & 
\colhead{(mag)} & 
\colhead{(mag)} & 
\colhead{(mag)} & 
\colhead{(mag)} 
}
\startdata
04180$-$3826 &  20048 & K2V &   26.8 &   25.0  &    0.8 &    0.7 &     8.89 &     1.11 & \ldots & \ldots &     0.50 &     0.14 \\
05229$-$4219 &  25148 & G5V &   15.1 &   15.1* &    1.1 &    0.8 &     8.70 &     0.76 & \ldots & \ldots &     1.22 &     0.38 \\
05542$-$2909 &  27901 & F3V &   18.5 &   17.2* &    1.5 &    1.2 &     6.34 &     0.43 &     1.49 &     0.18 &     1.06 & \ldots \\
06018$-$7158 &  28573 & K2/K3IV &   15.0 &   15.6  &    0.9 &    0.8 &     9.77 &     1.09 & \ldots & \ldots &     0.60 &     0.12 \\
06032$+$1922 &  28671 & G0V &   16.8 &   18.0  &    0.9 &    0.6 &     9.28 &     0.71 & \ldots & \ldots &     1.90 &     0.48 \\
08369$-$7857 & \ldots & K4Ve & \ldots &   11.5  &    0.8 &    0.8 &    10.46 &     1.45 & \ldots & \ldots &     0.39 &     0.21 \\
08398$-$6604 &  42496 & K1V: &   16.6 &   17.7* &    0.9 &    0.7 &     9.73 &     0.97 &     1.93 &     0.73 &     1.51 &     0.34 \\
08494$+$0341 &  43304 & G5 &   28.2 &   29.7  &    1.0 &    0.5 &     7.96 &     0.79 & \ldots & \ldots &     2.94 &     0.52 \\
09149$+$0427 &  45383 & K0 &   55.7 &   46.4  &    0.8 &    0.5 &     7.91 &     1.19 &     3.98 & \ldots &     2.13 &     0.46 \\
09186$+$2049 &  45671 & F5 &    9.2 &    9.1  &    1.3 &    1.1 &     8.64 &     0.51 & \ldots & \ldots &     0.77 & \ldots \\
09275$-$5806 &  46388 & G2V &   20.0 &   19.5* &    1.2 &    1.1 &     7.20 &     0.67 &     0.73 &     0.34 & \ldots & \ldots \\
09320$-$0111 &  46776 & A3V &    6.3 &    5.4* &    3.6 &    3.3 &     4.54 &     0.16 &     0.44 &     0.16 & \ldots & \ldots \\
09466$-$4955 &  47969 & G5IV &    6.8 &    6.9  &    1.4 &    1.2 &     8.73 &     0.83 &     0.99 &     0.10 &     1.23 &     0.17 \\
09510$-$8004 &  48320 & B8IV &    3.9 &    6.3  &    2.4 &    1.6 &     6.47 &     0.12 &     2.05 &     0.14 &     1.69 &     0.18 \\
10112$-$3245 &  49883 & G1V &   11.6 &   11.9  &    1.3 &    0.8 &     8.30 &     0.68 &     3.04 &     0.87 &     2.14 &     0.48 \\
11102$-$1122 &  54580 & G0 &   16.2 &   16.4* &    1.1 &    1.1 &     7.70 &     0.72 &     0.30 &     0.14 &     0.22 &     0.19 \\
11238$-$3829 &  55628 & K1V &   20.7 &   20.3* &    0.9 &    0.9 &     8.51 &     0.83 &     0.37 &     0.17 &     0.39 &     0.46 \\
11425$+$2355 &  57112 & G5 &   21.9 &   23.3* &    0.9 &    0.6 &     9.02 &     0.83 & \ldots & \ldots &     1.51 &     0.07 \\
11525$-$1408 &  57894 & G0V &   15.6 &   17.5* &    1.2 &    1.0 &     7.66 &     0.64 &     1.30 &     0.19 &     1.15 &     0.01 \\
13072$-$5420 &  64025 & G1/G2V &    1.4 &    4.8  &    1.5 &    1.2 &     9.08 &     0.73 & \ldots & \ldots &     0.90 &     0.06 \\
13081$-$7719 &  64091 & K3V &   23.3 &   26.1  &    0.9 &    0.6 &     8.76 &     1.01 & \ldots & \ldots &     1.79 &     0.07 \\
13306$-$4238 &  65906 & K4V &   30.6 &   31.2  &    0.7 &    0.6 &     9.56 &     1.47 & \ldots & \ldots &     0.97 &     0.11 \\
13377$-$2337 & \ldots & G5 & \ldots &   12.3  &    0.9 &    0.7 &    10.17 &     0.80 &     2.10 & \ldots &     0.25 & \ldots \\
13598$-$0333 &  68380 & F8V &   12.9 &   11.0* &    1.8 &    1.5 &     6.36 &     0.56 &     1.13 &     0.12 &     0.93 & \ldots \\
14020$-$2108 &  68552 & K0/K1V &   20.0 &   22.1* &    0.8 &    0.6 &     9.51 &     0.97 &     2.77 &     0.06 &     1.64 &     0.18 \\
14383$-$4954 &  71577 & A0IV &    4.2 &    4.3* &    2.7 &    2.2 &     6.51 &     0.05 &     0.98 &     0.20 & \ldots & \ldots \\
14453$-$3609 &  72140 & A1IV/V &    4.8 &    5.1  &    2.1 &    1.6 &     7.37 &     0.12 &     1.44 &     0.23 & \ldots & \ldots \\
14589$+$0636 &  73314 & K0? &   23.2 &   25.5* &    0.9 &    0.8 &     8.45 &     0.88 &     0.69 &     0.12 &     0.56 &     0.11 \\
15035$-$4035 &  73667 & F3V &    6.1 &    5.7  &    1.4 &    1.3 &     8.82 &     0.54 &     0.61 & \ldots &     0.44 & \ldots \\
15042$-$1530 &  73724 & F5/F6V &    8.4 &    7.9  &    1.5 &    1.2 &     8.03 &     0.60 &     1.15 &     0.51 & \ldots & \ldots \\
15251$-$2340 &  75478 & G2V &   15.3 &   15.5  &    1.0 &    1.0 &     8.48 &     0.65 &     0.52 &     0.36 &     0.28 &     0.14 \\
15282$-$0921 &  75718 & K2V &   48.6 &   51.4* &    0.9 &    0.6 &     6.89 &     0.86 &     3.40 &     0.04 &     2.21 &     0.24 \\
15290$-$2852 &  75790 & G0V &   17.1 &   14.6  &    1.6 &    1.3 &     6.43 &     0.68 &     0.93 & \ldots &     0.90 &     0.02 \\
15317$+$0053 &  76031 & G0 &   19.7 &   23.8* &    1.1 &    0.9 &     7.42 &     0.71 &     1.56 &     0.43 &     1.25 &     0.15 \\
15332$-$2429 &  76143 & A7+k... &   10.3 &    9.5* &    2.0 &    1.4 &     6.26 &     0.38 &     1.75 &     0.17 &     1.18 &     0.02 \\
15348$-$2808 &  76275 & G8V &   17.4 &   17.4  &    1.0 &    0.7 &     9.08 &     0.77 &     2.46 &     0.16 &     1.84 &     0.16 \\
15548$-$6554 &  77921 & G2V &   21.0 &   22.1* &    1.1 &    1.0 &     7.55 &     0.69 &     0.52 &     0.61 &     0.35 & \ldots \\
16544$-$3806 &  82709 & G0V &   15.6 &   17.3  &    1.1 &    0.7 &     8.40 &     0.69 &     3.27 &     0.15 &     2.32 & \ldots \\
17018$-$5108 &  83321 & A7V &    9.6 &    5.8  &    2.3 &    2.1 &     6.43 &     0.32 &     0.47 &     0.35 & \ldots & \ldots \\
17066$+$0039 &  83716 & K0V &   17.3 &   18.4* &    0.8 &    0.8 &     9.81 &     1.00 &     0.05 &     0.07 &     0.07 &     0.06 \\
17221$-$7007 &  84979 & B8/B9Vn... &    2.5 &    2.6* &    4.2 &    4.1 &     5.39 & \ldots &     0.15 &     0.22 & \ldots & \ldots \\
17575$-$5740 &  87914 & K0V &   25.1 &   24.9* &    0.8 &    0.7 &     9.31 &     1.00 &     0.42 &     0.22 &     0.28 &     0.14 \\
17584$+$0428 &  87991 & K8 &   22.9 &   26.3* &    0.7 &    0.6 &     9.80 &     1.40 &     1.16 & \ldots &     0.48 & \ldots \\
18092$-$2211 &  88932 & K3/K4V &   30.5 &   28.2* &    0.8 &    0.7 &     8.88 &     1.04 &     0.79 &     0.39 &     0.14 &     0.17 \\
18150$-$5018 & \ldots & A0V & \ldots &    6.2  &    1.9 &    1.7 &     7.22 & \ldots &     0.79 &     0.51 &     1.36 & \ldots \\
18368$-$2617 &  91253 & G3V &   21.5 &   18.7* &    1.1 &    1.0 &     7.80 &     0.70 &     0.78 &     0.25 &     0.43 &     0.20 \\
18480$-$1009 &  92250 & K0 &   36.0 &   41.9* &    0.8 &    0.5 &     8.45 &     0.93 &     3.65 & \ldots &     2.18 &     0.02 \\
19581$-$4808 &  98274 & G0V &   18.6 &   18.0* &    1.1 &    0.8 &     8.35 &     0.68 &     2.28 &     0.62 &     1.68 &     0.18 \\
20057$-$3743 &  98979 & F3V &    6.7 &    9.6  &    1.4 &    1.1 &     7.94 &     0.49 &     1.20 & \ldots &     0.41 &     0.06 \\
20104$-$1923 &  99391 & G0V &   16.4 &   14.6  &    1.4 &    0.8 &     7.28 &     0.65 &     3.09 &     0.38 &     2.29 &     0.17 \\
20202$-$3435 & 100266 & F8V &   16.1 &   15.9* &    1.5 &    1.2 &     6.62 &     0.66 &     1.24 &     0.09 &     1.10 &     0.02 \\
20217$-$3637 & 100417 & G0/G1V &   16.8 &   15.4* &    1.2 &    1.1 &     7.49 &     0.66 &     0.49 &     0.15 &     0.20 &     0.05 \\
20248$-$1943 & 100685 & K3V &   18.3 &   17.0  &    0.9 &    0.8 &     9.40 &     0.98 & \ldots & \ldots &     0.68 &     0.11 \\
21073$-$5702 & 104256 & K1V &   18.6 &   20.8* &    0.9 &    0.7 &     8.82 &     0.84 &     2.33 &     0.13 &     1.58 &     0.05 \\
22116$-$3428 & 109561 & K1/K2III+.. &    4.2 &    4.2* &    2.7 &    1.9 &     6.68 &     0.64 &     1.67 &     0.63 & \ldots & \ldots 
\enddata
\end{deluxetable}

\end{document}